\documentclass[a4paper,12pt]{article} 
\pdfoutput=1
\usepackage{graphicx,rotating,hyperref,slashed,amsmath,xcolor,amssymb,amsfonts,colortbl,cite} 
\makeatletter
\hypersetup{colorlinks,bookmarksopen,bookmarksnumbered,
linkcolor=blus,pdfstartview=FitH,urlcolor=rossos,citecolor=verde} 
\allowdisplaybreaks	  
\numberwithin{equation}{section}

\hypersetup{%
    ,urlcolor=black
    ,citecolor=black
    ,linkcolor=black
    }
\usepackage{mciteplus}

\AtBeginDocument{
  \hypersetup{
    urlcolor=blue,
    citecolor=blue,
    linkcolor=black,
  }%
}

\def\beq{\begin{equation}}
\def\eeq{\end{equation}}

\renewcommand\[{\left[}
\renewcommand\]{\right]}

\newcommand\ees{\end{eqnarray}}
\newcommand\bees{\begin{eqnarray}}

\def\bea{\begin{eqnarray}}
\def\eea{\end{eqnarray}}

\def\0{{\boldsymbol 0}}

\def\lsim{\mathrel{\rlap{\lower3pt\hbox{\hskip0pt$\sim$}}
   \raise1pt\hbox{$<$}}}         
\def\gsim{\mathrel{\rlap{\lower4pt\hbox{\hskip1pt$\sim$}}
   \raise1pt\hbox{$>$}}}         

 \newcommand{\sfootnote}[1]{}
\definecolor{bluc}{cmyk}{1,1,0,0.1}
\definecolor{rossoCP3}{cmyk}{0,.88,.77,.40}
\definecolor{rosso}{cmyk}{0,1,1,0.4}
\definecolor{rossos}{cmyk}{0,1,1,0.55}
\definecolor{rossoc}{cmyk}{0,1,1,0.2}
\definecolor{verdes}{cmyk}{0.92,0,0.59,0.4}

\hypersetup{colorlinks, bookmarksopen, bookmarksnumbered,
citecolor=verdes, linkcolor=bluc, pdfstartview=FitH, urlcolor=rossos}

\usepackage{multicol}
\usepackage{color}
\definecolor{rosso}{cmyk}{0,1,1,0.4}
\definecolor{rossos}{cmyk}{0,1,1,0.55}
\definecolor{rossoc}{cmyk}{0,1,1,0.2}
\definecolor{blu}{cmyk}{1,1,0,0.3}
\definecolor{blus}{cmyk}{1,1,0,0.6}
\definecolor{bluc}{cmyk}{1,1,0,0.1}
\definecolor{verde}{cmyk}{0.92,0,0.59,0.25}
\definecolor{verdec}{cmyk}{0.92,0,0.59,0.15}
\definecolor{verdes}{cmyk}{0.92,0,0.59,0.4}

\skip\@mpfootins = \skip\footins
\renewcommand\&{&}

\def\circa#1{\,\raise.3ex\hbox{$#1$\kern-.75em\lower1ex\hbox{$\sim$}}\,}

\newcommand{\be}{\begin{equation}}
\newcommand{\ee}{\end{equation}}
\newfam\rsfsfam
\def\mathscr#1{{\fam\rsfsfam\relax#1}}

\def\circa#1{\,\raise.3ex\hbox{$#1$\kern-.75em\lower1ex\hbox{$\sim$}}\,}
\makeatletter

\def\hhref#1{\href{http://arxiv.org/abs/#1}{arXiv:#1}} 

\newcommand{\doi}[1]{\href{http://dx.doi.org/#1}{[doi]}}

\def\hhref#1{\href{http://arxiv.org/abs/#1}{arXiv:#1}} 
 
\def\art{\@ifnextchar[{\eart}{\oart}}
\def\eart[#1]#2#3#4#5#6{{\rm #2}, {\em #3 \bf #4} {\rm (#6) #5} ({\em #1})}

\def\article{\@ifnextchar[{\earticle}{\oarticle}}
\def\oarticle#1#2#3#4#5#6{{\rm #1}, {\em ``#6''}, {\rm #2 #3 (#5) #4}}
\def\earticle[#1]#2#3#4#5#6#7{{\rm #2}, {\em ``#7''}, {\rm #3 #4 (#6) #5}  [\hhref{#1}]}
\def\hepart[#1]#2{{\rm #2, \em#1}}
\def\heparticle[#1]#2#3{#2, {\em ``#3''} [\hhref{#1}]}

\newcounter{alphaequation}[equation]
\def\thealphaequation{\theequation\hbox to
0.6em{\hfil\alph{alphaequation}\hfil}}
\def\eqnsystem#1{
\def\@eqnnum{{\rm (\thealphaequation)}}
\def\@@eqncr{\let\@tempa\relax \ifcase\@eqcnt \def\@tempa{& & &} \or
  \def\@tempa{& &}\or \def\@tempa{&}\fi\@tempa
  \if@eqnsw\@eqnnum\refstepcounter{alphaequation}\fi
\global\@eqnswtrue\global\@eqcnt=0\cr}
\refstepcounter{equation} \let\@currentlabel\theequation \def\@tempb{#1}
\ifx\@tempb\empty\else\label{#1}\fi
\refstepcounter{alphaequation}
\let\@currentlabel\thealphaequation
\global\@eqnswtrue\global\@eqcnt=0 \tabskip\@centering\let\\=\@eqncr
$$\halign to \displaywidth\bgroup \@eqnsel\hskip\@centering
$\displaystyle\tabskip\z@{##}$&\global\@eqcnt\@ne
\hskip2\arraycolsep\hfil${##}$\hfil& \global\@eqcnt\tw@\hskip2\arraycolsep
$\displaystyle\tabskip\z@{##}$\hfil
\tabskip\@centering&\llap{##}\tabskip\z@\cr}
\def\endeqnsystem{\@@eqncr\egroup$$\global\@ignoretrue} \makeatother

\definecolor{fiorentina}{rgb}{.5,0,.5}

\begin{document}

\setcounter{page}{1} \baselineskip=15.5pt \thispagestyle{empty}

\vspace{0.8cm}
\begin{center}

{\fontsize{19}{28}\selectfont  \sffamily \bfseries {Hidden conformal symmetries for black holes
\\ \vskip0.2cm
  in modified gravity}}

\end{center}

\vspace{0.2cm}

\begin{center}
{\fontsize{13}{30}\selectfont  Bill Atkins$^{1}$, Gianmassimo Tasinato$^{1,2}$ } 
\end{center}

\begin{center}
{\small
\vskip 8pt
\textsl{$^{1}$ Physics Department, Swansea University, SA28PP, United Kingdom}\\
\textsl{$^{2}$ Dipartimento di Fisica e Astronomia, Universit\`a di Bologna, and }\\
\textsl{INFN, Sezione di Bologna, I.S. FLAG, viale B. Pichat 6/2, 40127 Bologna, Italia}\\
\vskip0.1cm
\textsl{\texttt{email}: bill.atkins847 at gmail.com, \,g.tasinato2208 at gmail.com }\\
\vskip 7pt
}
\end{center}

\smallskip
\begin{abstract}
\noindent
We determine hidden conformal symmetries behind the evolution equations of black hole perturbations in a vector-tensor theory of  gravity. Such hidden symmetries are valid everywhere in the exterior region of a spherically symmetric, asymptotically flat black hole geometry. They allow us to factorize second order operators controlling the black hole perturbations into a product of two commuting first order operators. As a consequence, we are able to analytically determine the most general time-dependent solutions for the black hole perturbation equations.  We focus on solutions belonging to a highest weight representation of a conformal symmetry, showing that they correspond to quasi-bound states with an ingoing behaviour into the black hole horizon, and  exponential decay at spatial infinity. Their time-dependence is characterized by purely imaginary frequencies, with imaginary parts separated by integer numbers, as the overtones of quasi-normal modes in General Relativity. 
\end{abstract}

\newpage
\section{Introduction}

The study of emergent   conformal symmetries is a topic of active research for black holes in General Relativity. Conformal symmetries are important
for understanding black hole entropy in terms
of microstate counting for BTZ  \cite{Strominger:1997eq} and extremal \cite{Guica:2008mu} black holes, as well as  for configurations equipped with anti-de Sitter symmetries  (see e.g. \cite{Bardeen:1999px,Govindarajan:2000ag,Birmingham:2001qa,Hartman:2008pb}). Given
the need of a deeper understanding of black hole entropy to address the black hole
information problem,   exploring 
the emergence of conformal symmetries in black hole geometries may prove particularly insightful. 

Conformal symmetries emerge in the near horizon region of Schwarzschild black
holes. They are associated with diffeomorphism invariance of geometrical
quantities in the proximity of the black hole horizon \cite{Carlip:1998wz,Solodukhin:1998tc}, or with the properties of the near horizon
optical metric \cite{Sachs:2001qb}. However, they can also appear as {\it hidden symmetries} \cite{Bertini:2011ga}, without apparent relation to any geometric properties of the black hole space-time (see also \cite{Chen:2010ik}). This is reminiscent of the situation for Kerr \cite{Castro:2010fd}, whereby such hidden symmetries and their Virasoro central charge extensions have  been related with the black hole entropy. The interesting results of said hidden symmetries in \cite{Bertini:2011ga} have been further explored in the literature: see (for example), \cite{Cvetic:2011hp,Franzin:2011wi,Virmani:2012kw,Kim:2012mh,Ghezelbash:2012qn}. Recently, they have played a key role in  \cite{Charalambous:2022rre} where they have been used to uncover a black hole  {\it Love symmetry}, able
to explain the vanishing of Love numbers for four dimensional black holes in terms of underlying symmetry structures. Further work has demonstrated that conformal symmetries emerge not only near the horizon, but also in the proximity of the photon ring region \cite{EventHorizonTelescope:2019dse,Gralla:2019xty,Johnson:2019ljv,Himwich:2020msm,Gralla:2020yvo,Gralla:2020srx,Li:2021zct,Raffaelli:2021gzh,Hadar:2022xag,Kapec:2022dvc}   
  of a black hole, explaining some of its features.

It is interesting to explore if  these results pertain to General Relativity, or instead whether hidden conformal  symmetries can be found in  alternative theories of gravity (see e.g. \cite{Barack:2018yly} for a comprehensive review).
In this work we focus on a specific case of   a vector-tensor theory of  gravity \cite{Tasinato:2014eka,Heisenberg:2014rta}, in which vector fields are non-minimally coupled with curvature.
 The dynamics of  parity-odd  fluctuations around a class of  spherically symmetric, asymptotically flat geometries  appear to be particularly simple.   We concentrate on a background solution with a geometry corresponding to a Schwarzschild black hole \cite{Chagoya:2016aar,Tasinato:2022vop}. The evolution
equations for vector and metric fluctuations are characterized by 
hidden conformal symmetries, associated with $SL(2,R)$ algebras, and their
extensions to centerless Virasoro symmetries.  This is somehow unexpected, since our geometrical configurations do not enjoy AdS asymptotics, nor correspond to extremal black hole geometries.  Interestingly,  our symmetries apply on  the {\it entire} exterior geometry of the configuration as opposed to only being realised in the near horizon limit.

The evolution equations for the vector-field and metric perturbations can be factorized into a product of two commuting first order operators. This property allows us to analytically determine the most general time-dependent solutions  for the system of equations.  In fact, we associate
this factorizability property with  emergent conformal symmetries behind the structure
of the equations controlling the  field fluctuations, and directly link the two structures.

We proceed to study the highest weight representation for one of the  $SL(2,R)$ algebras, and analytically determine the expression for the highest weight   time-dependent solutions, and their  descendants. As pointed out more generally in \cite{Chen:2010ik}, the highest weight representations  are known to have common properties with quasi-normal
modes of black holes in General Relativity. Indeed, we find that elements belonging to the highest weight multiplets have frequencies separated by   integer numbers, resembling the behaviour of overtones of black hole quasi-normal modes. We note, however, that although our  solutions have the time-dependent profile of ingoing modes into the black hole horizon, they do not describe   outgoing modes asymptotically far from it, as quasi-normal modes do. Instead, they  decay exponentially with the radial distance from the black hole horizon, behaving as quasi-bound
states.

The details of the system and the dynamics of fluctuations around a spherically symmetric
Schwarzschild solution are discussed in sections \ref{sec_setup} and \ref{sec_fluct}, where
we analyze in detail the emergent conformal symmetries. Their physical implications
are discussed in section \ref{sec_pheno}. Our conclusions and further considerations may be found in section \ref{sec_out}.

\section{System under consideration}
\label{sec_setup}

We consider Einstein-Maxwell gravity, including a non-minimal coupling 
between a vector field $A_\mu$ and the Einstein tensor $G_{\mu\nu}$. The Lagrangian
density is  \cite{Tasinato:2014eka,Heisenberg:2014rta}
\be
\label{starLAG}
{\cal L}\,=\,\frac12 {R}-\frac14\, F_{\mu\nu}  F^{\mu\nu} 
 +\frac{1}{4}\,G_{\mu\nu} \,A^\mu A^\nu\,.
\ee
This system leads to second order equations of motion, hence it is 
free of Ostrogradsky instabilities. The choice of the non-minimal coupling  between
the vector and the curvature -- the factor $1/4$ in front of the $G_{\mu\nu} \,A^\mu A^\nu$ combination -- 
leads to particularly simple black hole and spherically symmetric solutions, see e.g. \cite{Chagoya:2016aar,Tasinato:2022vop} (see also \cite{Minamitsuji:2016ydr,Cisterna:2016nwq,Babichev:2017rti,Chagoya:2017fyl,Heisenberg:2017xda,Chagoya:2017ojn,Filippini:2017kov} for further
developments on this topic), and we focus on this specific choice in what follows. Notice that the theory breaks
a $U(1)$ Abelian symmetry through the direct coupling of the vector field
to gravity. 
We
 consider the spherically symmetric Ansatz
 \bea
 \label{metrans1}
 d s^2\,=\,\bar g_{\mu\nu}\,d x^\mu d x^\nu&=&-\bar A(r)\,d t^2+\frac{d r^2}{\bar B(r)}\,+r^2\, d\theta^2+r^2\,\sin^2{\theta}\, d\phi^2\,,
\eea
for the metric,
and an electric-type  Ansatz
\bea
 \label{vecans1}
\bar V_\mu d x^\mu
&=&\bar \alpha_0(r)\,d t+\bar \Pi(r)\,d r\,,
 \eea
 for the vector field. From now on,  a bar denotes background quantities
 that depend on the radial coordinate only.  Since the Abelian gauge symmetry is broken, we can not gauge away the radial
 component of the vector profile. 
 
 The corresponding
  Einstein and vector field equations admit two branches of solutions (we refer
  the reader  to  \cite{Chagoya:2016aar} for details on the system of equations). One branch contains vanishing vector radial component $\bar{\Pi}(r)=0$, and is continuously connected
  with the Reissner-Nordstr\"om configuration.  No known exact solutions exist in this branch.  The equations of motion for the other branch -- on which we focus
  our attention --  are satisfied by choosing profiles for $\bar B(r)$, $\bar \alpha_0(r)$, and $\bar  \Pi(r)$
  obeying the following relations
  \bea
\bar B(r)&=&\frac{\bar A(r)}{\bar A(r)+r\,\bar A'(r)}
\,,
\label{meqG}
\\
\left[\frac{d}{dr} \left( \frac{r \,\bar \alpha_0(r)}{2} \right)\right]^2&=&
\frac{d}{dr} \left( {r \,\bar A(r)} \right)\,,
\label{meqA}
\\
\bar \Pi(r)&=&
\sqrt{\frac{\bar \alpha_0^2(r)-4 \bar A(r) }{\bar A(r)\,\bar B (r)}}
\,.
\label{meqPI}
\eea
Hence the configuration is determined by a choice of the arbitrary function $\bar A(r)$, compatible
with the boundary conditions one wishes to impose. For example, choosing $\bar A(r)\,=\,1-2M/r$,
 one finds 
   an asymptotically flat, spherically symmetric solution corresponding to a Schwarzschild geometry
  \bea
d s^2&=&-\left( 1-\frac{2M}{r}\right) d t^2+\frac{d r^2}{\left( 1-\frac{2M}{r}\right)}
+r^2\,d \Omega^2\,,
\label{extgeo1}
\\
\bar \alpha_0(r)&=&2+\frac{2 Q}{r}\,,
\\
\label{extPI}
\bar \Pi(r)&=&\frac{2 \sqrt{ Q^2+2( M+Q) r}}{r-2M}\,,
\eea
characterized by 
 two arbitrary constants $M$, $Q$. Notice that -- differently from the Reissner-Nordstr\"om
 black hole -- 
 the background geometry does not depend on $Q$ (whilst, as we will learn, 
 the dynamics of fluctuations  depends on this quantity).
  This geometrical configuration
 is equipped with a Schwarzschild horizon  located at radius $R_S\,=\,2 M$.

 \smallskip 
 
 However a solution
 with a Schwarzschild horizon  is not 
 the unique asymptotically flat configuration solving equations \eqref{meqG}-\eqref{meqPI}. There is the possibility to  smoothly  connect the exterior Schwarzschild  geometry \eqref{extgeo1} to a regular interior
 configuration with no horizon, dubbed `ultracompact vector star' in  \cite{Tasinato:2022vop}.
We briefly discuss this configuration since, in an appropriate limit, its symmetry
  properties  resemble what we  will find next for the dynamics
  of the fluctuations. The metric and vector-field components for the interior solution -- recall
  our metric Ansatz \eqref{metrans1} --
  are
\bea
\label{solA2}
\bar A(r) &=&\sigma^2 +\frac{2 \sigma (1-\sigma)}{1+\gamma}
\,\left( \frac{r}{\bar R}\right)^\gamma+
\frac{  (1-\sigma)^2}{1+2\gamma}
\,\left( \frac{r}{\bar R}\right)^{2\gamma}
\\
\label{solB2}
\bar B(r) &=&\frac{\bar A}{\left[ \sigma+(1-\sigma) (r/\bar R)^{\gamma}\right]^2}
\eea
and
\bea
\bar \alpha_0(r) &=&2\left(Q+\frac{\gamma (1-\sigma)}{1+\gamma}
\right) \frac{\bar R}{r}+2 \sigma+\frac{2(1-\sigma)}{1+\gamma}
\,\left( \frac{r}{\bar R}\right)^{\gamma}\,,
\label{solPI2}
\eea
while the (long) expression for $\bar \Pi(r)$ can be found by plugging the previous equations
into equation \eqref{meqPI}.
This solution is
  characterized by the additional dimensionless constants $0\le\sigma\le1$ and $\gamma\ge0$, as well as 
  a  radius $\bar R$ 
  corresponding to the boundary of the vector star
  (the $Q$ is the same as in the exterior solution above). 
  This configuration solves the equations
  in Appendix \ref{appA_ev} under conditions \eqref{meqG}-\eqref{meqPI}
  which characterise  the branch of solutions
  we are interested in. It
  connects smoothly with the exterior configuration, with no need to consider contributions from the extra surface 
  energy momentum tensor, if its parameters satisfy the relation
   \be
\bar   R\,=\,\frac{(1+\gamma) (1+2\gamma)}{2\gamma (1-\sigma) (1+\gamma+\sigma
\gamma)}\,R_S\,\ge\,R_S
 \,.
  \ee
 The solutions described by equation \eqref{extgeo1} (in the exterior) and equation \eqref{solA2} (in the interior)
  corresponds to an asymptotically flat vector star, with 
  compactness
  \be
{\cal C}\,\equiv\,\frac{M}{\bar R}\,=\,
\frac{\gamma (1-\sigma) (1+\gamma+\sigma
\gamma)}{(1+\gamma) (1+2\gamma)}\,\le\,\frac12
\ee
smaller than a Schwarzschild black hole. The black hole compactness limit ${\cal C}=1/2$
is reached for $\sigma\to0$ and $\gamma\to\infty$. These objects
can avoid Buchdahl theorem and be as compact as black holes, thanks to their internal anisotropic stress \cite{Tasinato:2022vop}.

\smallskip

When  $\sigma\to0$,
the configuration of equations \eqref{solA2} and \eqref{solB2} becomes singular and 
 develops  self-similar properties, with a scaling symmetry resembling
 a singular isothermal sphere \cite{Tasinato:2022vop}.
  This feature  indicates
 that scaling symmetries and conformal transformations can play an important role in the characterization of this system. In fact, a
 rich pattern of conformal symmetries emerge  when studying the dynamics of parity-odd fluctuations, as
 we are going to discuss in the following.

\section{Parity-odd fluctuations and conformal symmetries}
\label{sec_fluct}

We now  consider the dynamics of parity odd fluctuations around a spherically symmetric
background  configuration satisfying the
system of equations \eqref{meqG}-\eqref{meqPI}. For definiteness,
we  focus on the exterior Schwarzschild geometry described in equations \eqref{extgeo1}-\eqref{extPI}.
We demonstrate that 
  the evolution 
equations for the fluctuations   enjoy a large set of  symmetries (including
conformal $SL(2,R)$ symmetries), which allows us to analytically characterize
their time-dependent solutions,  and their  corresponding properties.


Fluctuations around our background configuration are parameterized in terms of small quantities $h_{\mu\nu}$
and $a_\mu$, as
\bea
g_{\mu\nu}&=&\bar g_{\mu\nu}+h_{\mu\nu}\,,
\\
A_\mu&=&\bar A_\mu+a_{\mu}\,.
\eea
In Regge-Wheeler gauge,
the metric fluctuations 
 are controlled by  
two non-vanishing
components $h_0$ and $h_1$, which depend on time and on the radial direction:
\be \small{
h_{\mu\nu}=\begin{pmatrix}
0 & 0 & - h_0(t,r) (\partial_{\varphi}/\sin \theta) & h_0(t,r) \sin \theta \,\partial_{\theta} \\
 0 & 0 & -h_1(t,r)\,(\partial_{\varphi}/\sin \theta)& h_1(t,r) \sin \theta \,
   \partial_{\theta} \\
 - h_0(t,r) (\partial_{\varphi}/\sin \theta) & -h_1(t,r)\,(\partial_{\varphi}/\sin \theta) &
0
& 0 
\\
 h_0(t,r) \sin \theta \, \partial_{\theta} &
   h_1(t,r) \sin \theta \,\partial_{\theta}& 0
   &
   0
   \\
\end{pmatrix} } Y(\theta,\varphi).
\ee
with $Y(\theta,\varphi)$ denoting  spherical harmonics.
There is a single parity-odd component for the 
perturbations $a_{\mu}$ of the vector field, which we denote with $\beta(t,r)$:
 \be
\hspace*{-1cm}
a_{\mu}=\left\{0
 ,0
 , \,\frac{r\,\bar \alpha_0^2(r)}{4}\,\beta(t,r) \sin \theta^{-1}\, \partial_{\varphi} 
 ,
 - \frac{r\,\bar \alpha_0^2(r)}{4}\,\beta(t,r) \sin \theta \, \partial_{\theta}\right\}Y(\theta,\varphi).
\ee
 For convenience, we multiply the parity-odd vector fluctuation $\beta(t,r)$
by the background quantity ${r\,\bar \alpha_0^2(r)}/{4}$ in order to simplify the corresponding evolution
equations. 
 It is straightforward to determine the quadratic action governing the dynamics of fluctuations $h_0(t,r)$, $h_1(t,r)$, and $\beta(t,r)$ around a configuration satisfying equations  \eqref{meqG}-\eqref{meqPI}. This task
 has been carried out~\footnote{The work  \cite{Kase:2018voo}  shows that the dynamics of parity odd fluctuations
 around spherically symmetric solutions for the theory \eqref{starLAG} are generically
 plagued by instabilities in the near-horizon region. 
 However, 
  the specific background configuration \eqref{extgeo1}-\eqref{extPI} we are considering avoids their arguments. In fact -- in the notation of \cite{Kase:2018voo} -- it leads a vanishing quantity $C_1$ and $C_6$ as defined in Appendix C of  \cite{Kase:2018voo}. But  $C_1$ is assumed
 to be non-vanishing in \cite{Kase:2018voo},  since it appears in the denominator of many equations, upon
 solving constraint equations. Hence, the instability arguments as developed in \cite{Kase:2018voo}
 do not apply in the present instance.}
  in \cite{Kase:2018voo}.  
 We write
 the corresponding equations of motion  in Appendix \ref{appA_ev}. Upon using
 conditions \eqref{meqG}-\eqref{meqPI}, formulas simplify dramatically.
 We can algebraically solve for $h_1(t,r)$, and express this quantity as a linear
 combination of $h_0(t,r)$ and $\beta(t,r)$ in the exterior geometry of equations \eqref{extgeo1}-\eqref{extPI}:
 \bea
 h_1&=&
 \frac{r}{r+Q}\frac{\sqrt{Q^2+2 (M+Q)r}}{r-2M}\,h_0-\frac{\sqrt{Q^2+2 (M+Q)r}}{2}\,\beta
 \nonumber
 \\
 &-&\frac{r^2}{2(\ell+2)(\ell-1)}\,\partial_t \beta+\frac{r^2 (r+Q)}{2(\ell+2)(\ell-1)}\,\partial_t \partial_r \beta
 -\frac{r^3\,\sqrt{Q^2+2 (M+Q)r}}{2(r-2M)(\ell+2) (\ell-1)}\,\partial_t^2\beta\,.
 \nonumber\\
 \label{condH1}
 \eea
 The quantity $\ell$
 is the multipole number, an integer $\ell\ge2$. 
  Hence,   only two degrees of freedom are dynamical in this system: we take them to be $\beta(t,r)$ and $h_0(t,r)$. 
 Upon
 substituting condition \eqref{condH1} into the remaining equations, the dynamics of the vector fluctuations $\beta(t,r)$
 are decoupled from the metric perturbation $h_0(t,r)$. We study the interesting
 symmetry
 properties of the decoupled vector equation in section \ref{sec_vec}, and then
 analyze the metric fluctuation dynamics in section \ref{sec_met}. The physical
 implications of our findings are studied in section \ref{sec_pheno}.
  
\subsection{Vector-field perturbations and associated symmetries}
\label{sec_vec}

After inserting equation \eqref{condH1} into 
the remaining evolution equations, the vector fluctuation
$\beta(t,r)$   decouples from the metric fluctuation $h_0(t,r)$.
  The evolution equation reads
\bea
\label{veceq1}
{\cal E}_\ell\left[ \beta \right] &=&0\,.
\eea
  ${\cal E}_\ell[\dots]$ is a linear operator defined in the exterior
  black hole region, $r\ge2M$:
 \bea
{\cal E}_\ell\left[ \beta(t,r)\right] &=&
 2 \sqrt{\bar \Delta(r)}\, \partial_r \beta+\bar \Delta (r)  \partial_r^2 \beta(t,r)-2 \bar \Sigma (r) \partial_t \partial_r \beta 
\nonumber
\\&+&
\frac{\bar \Sigma^2 (r) \partial_t^2\beta}{\bar \Delta (r)}
-\bar \Sigma '(r)
\partial_t \beta
-\ell (\ell+1)\beta  \,,
 \label{veceq1a}
\eea
where
 the integer $\ell\ge2$  denotes the multipole number. 
The definition of the radial-dependent functions $\bar \Delta$ and $\bar \Sigma$ are
given in terms of the background quantities 
\bea
\bar \Delta(r)&=&\frac{r^2 \bar \alpha_0^2}{4 (\bar A+r  \bar A')} 
\label{dg1}
\,,
\\
&=&(r+Q)^2
\,,
\label{ds1}
\eea
and
\bea
\label{dsi1}
\bar  \Sigma(r)&=&\frac{r^2 \bar \alpha_0}{4 \bar A} \sqrt{\frac{\bar \alpha_0^2-4 \bar A}{
\bar
A+r \bar A'}}
\,,
\\
\label{dsi2}
&=&\frac{r (r+Q)}{r-2M}\,\sqrt{Q^2+r (2 M+2 Q)}\,.
\eea
In passing from the first to the second line in each of the previous
equations we are specializing to  the Schwarzschild-type `exterior' solution
of equations \eqref{extgeo1}-\eqref{extPI}.  

The vector equation \eqref{veceq1a}
  has  intriguing properties.  
Although it involves
second derivatives along the temporal and radial coordinates,
it can be factorized as a product of two commuting
  first-order operators $D^{+}$ and $D^{-}$,
  as
  
  \be
  \label{factpr1}
{\cal E}_\ell\left[ \beta(t,r)\right]\,=\,\left(D^{-}
   \,D^{+}\right)\left[ \beta(t,r)\right]\,=\,\left(D^{+}
   \,D^{-}\right)\left[ \beta(t,r)\right]
  \ee
  with
  \bea
D^{\pm}\left[ \beta(t,r)\right]&=&
\frac{\bar \Sigma}{\sqrt{\bar \Delta} }\,\partial_t \beta-\sqrt{\bar \Delta} \,\partial_r
 \beta-\sigma_{\pm} \beta\,.
 \label{def_lad}
\eea
The constants $\sigma_{\pm}$ in equation \eqref{def_lad} depend on the multipole number
as
\bea
\label{def_sig}
\sigma_+&=&1+\ell \; \; \; \; \,, \; \; \; \; \; 
\sigma_-\,=\,-\ell\,.
\eea
  Crucially, the factorizability feature \eqref{factpr1} is valid in the entire exterior
region of the black hole, and not only in the proximity of the horizon. Such a property
can be expected for the evolution of fluctuations in the context of extremal
black holes (for example, see 
 \cite{Berens:2022ebl}, their equation (2.30),   setting the extremal condition $r_S=2 r_Q$). However it is quite unexpected
 in our case, where we deal with a Schwarzschild background. 
The property \eqref{factpr1} of  the evolution equation  \eqref{veceq1}
  implies that, for any given multipole $\ell$,  the general solution $\beta(t,r)$ of the second order equation  is a linear combination of the  solutions 
   of the two first order equations
\bea
\label{redeq1}
D^{+}\left[ \beta(t,r)\right]&=&0\,,
\\
\label{redeq2}
D^{-}\left[ \beta(t,r)\right]&=&0\,.
\eea
The exact time-dependent solutions of   equations \eqref{redeq1} and \eqref{redeq2} are not difficult to determine.
The complete solution then reads
\be\label{full_sol}
\beta(t,r)\,=\,c_{+}\,\frac{F_{+}(t+r_\star)}{
(r+Q)^{\sigma_{+}}}+c_{-}\,\frac{F_{-}(t+r_\star)}{
(r+Q)^{\sigma_{-}}}\,,
\ee
where $c_{\pm}$ are two arbitrary constants, $\sigma_\pm$ appear as defined in equation \eqref{def_sig}, and $F_\pm$ are
two arbitrary functions of the combination
\bea
\label{def_tort}
t+r_\star\equiv t +\int^r\,\frac{\bar \Sigma(\tilde r)}{\bar \Delta(\tilde r)}\,d\tilde r
\,.
\eea
The `tortoise' radial coordinate $r_\star$ is given by the radial integral in equation \eqref{def_tort}: we will study
its properties in section \ref{sec_pheno}.
 Further
specifications of physically interesting solutions depend both on the boundary
conditions we are interested in, and on the structure
of the configurations we wish to study.

\smallskip
When focusing on static fluctuations,
first-order operators similar to our definition in equation \eqref{def_lad} have been introduced in \cite{Hui:2021vcv,Berens:2022ebl} as `ladder' operators in the context
of Schwarzschild solutions of General Relativity (see also \cite{Cardoso:2017qmj,BenAchour:2022uqo} for related
proposals).
 The ladder operators studied in \cite{Hui:2021vcv}  do not satisfy a  commutation
relation like our \eqref{factpr1}. Nevertheless they  allow one to generate
solutions of different multipole numbers starting from a given solution at multipole level $\ell$. Such features have proven
to be helpful for understanding the vanishing of Love
numbers of black hole solutions.

\smallskip

 We do not pursue this line of investigation. Instead, we analyse the full time-dependent solutions for the equations of motion, as given in equation \eqref{full_sol}. 
 We ask whether the particularly simple structure
 of our solutions
 \eqref{full_sol} can be associated with some underlying symmetry, possibly
 also explaining the factorizability property  \eqref{factpr1} of the equations 
 of motion. We  answer affirmatively, and we determine  underlying conformal
 symmetries behind the vector evolution equation. Interestingly, these conformal symmetries are not just a near-horizon
 property of the  configuration: they  extend to the entire exterior
 black hole geometry. 
 
 \subsubsection*{A first set of symmetries}

   We introduce the  first-order operators $L_p$ involving solely first derivatives
 \be
 \label{def_elp}   
L_p\,=\,-\frac{e^{\frac{p}{4 M}\left( t +r_\star\right)}}{\sqrt{\bar \Delta}}\left[\left(
{4 M \sqrt{\bar \Delta}+p \bar \Sigma} \right)\, \partial_t -p \,\bar \Delta\, \partial_r \right]\,,
\ee
where  $p$ is an arbitrary integer and the tortoise radial variable $r_\star$ is defined in equation \eqref{def_tort}. The
quantities $\bar \Delta$, ${\bar \Sigma}$ are given in equations \eqref{ds1} and \eqref{dsi1}.  
 The operators $L_p$ satisfy a centerless Virasoro algebra
\bea
\label{comm_ell}
 [ L_p, L_s ]\,=\,(p-s)\,L_{p+s}\,,
 \eea
 for every integer $p$, $s$. 
 Moreover, for each $p$, the $L_p$ operators commute with the operator ${\cal E}_\ell$ of \eqref{veceq1} controlling
 the vector equations of motion:
\be
[{\cal E}_\ell, L_p]\,=\,0\,,
\ee
 indicating that they can be used for generating solutions from  existing ones. Hence
 they represent a symmetry for the system. The choice of operators \eqref{def_elp} is inspired by the structure of operators studied in \cite{Bertini:2011ga}, adapted
 to the particular form of equations of motion in our context. In fact, 
 while a similar algebraic structure has been found in the near-horizon region of Schwarzschild
 black hole in GR \cite{Bertini:2011ga}, in our case the commutation relations are
 valid everywhere outside the horizon.
 
 \smallskip
 
 In fact, we can consider an $SL(2,R)$ conformal sub-algebra constituted
 by the three operators $L_{+1}$, $L_0$, $L_{-1}$, by focusing on 
 $|p|\le1$. The operator ${\cal E}_\ell$
 is associated with the quadratic Casimir of the $SL(2,R)$  algebra:
\be
 {\cal E}_\ell[\beta(t,r)]+\ell (\ell+1)\,\beta(t,r)
 \,=\,\left(
 L_0^2-\frac12 \left\{ L_1,L_{-1}\right\} \right)[\beta(t,r)]\,.
\ee
 Subsequently, solutions of the equations of motion can be related,  
 starting from a highest weight representation of the  $SL(2,R)$  algebra \cite{Chen:2010ik,Bertini:2011ga,Charalambous:2022rre}. 
   Such a representation
 is defined through the conditions
 \bea
 \label{high_con1}
 L_{1}\,\left[\beta^{(0)}(t,r)\right]&=&0\,,
 \\
  \label{high_con2}
 L_0\,\left[\beta^{(0)}(t,r)\right]&=&\sigma_{\pm} \,\beta^{(0)}(t,r)
\,.
 \eea
 Both the choices $\sigma_{\pm}$ are allowed in principle, depending
 on the physical structures one wishes to analyze, and also depending
 on whether one studies (as we do here) a highest weight or
 a lowest weight representation \cite{Charalambous:2022rre}. 
 It is not difficult to convince ourselves, using the commutation
 relations \eqref{comm_ell}, that
 a function $\beta^{(0)}(t,r)$  satisfying  the previous 
 conditions also satisfies the equations of motion ${\cal E}_\ell[\beta^{(0)}(t,r)] \,=\,0$. Moreover, starting from $\beta^{(0)}(t,r)$,  we can build its descendants as
 \be
 \label{def_desc}
 \beta^{(n)}(t,r)\,\equiv\,\left( L_{-1} \right)^{n}\,\left[\beta^{(0)}(t,r)\right]\,,
 \ee
 which also satisfy the equations of motion, and belong to the  conformal   multiplet associated with
  the highest weight solution $\beta^{(0)}$.
 
  The general
 solution of conditions \eqref{high_con1} and \eqref{high_con2} contains a plane-wave structure (but with a purely imaginary frequency). It results 
 \be
 \label{sol_bz}
 \beta^{(0)}(t,r)\,=\,c_+\,
 \frac{e^{-\frac{\sigma_+\,(t+r_\star)}{4 M}}}{ (r+Q)^{\sigma_+}}+c_- \,\frac{e^{-\frac{\sigma_-\,(t+r_\star)}{4 M}}}{ (r+Q)^{\sigma_-}}
 \,,
 \ee
for two arbitrary constants $c_{\pm}$. The solution \eqref{sol_bz}
 has the correct structure we determined in \eqref{full_sol}  for solving the vector equations of motion. As anticipated above, the exponential structure \eqref{sol_bz} indicates
 that the highest weight solution  \eqref{sol_bz} behaves as a plane wave
 with a purely imaginary frequency $\omega$ such that $4 i M \omega\,=\,\sigma_{\pm}$ for each of the two contributions
 in equation \eqref{sol_bz} (recall
 that the $\sigma_\pm$ of equation \eqref{def_sig} are integers).
  We will return in section \ref{sec_pheno} to study its properties and the properties of its descendants.

 \subsubsection*{A second set of symmetries}

 Interestingly, besides the  conformal symmetries associated
 with operators $L_p$ of equation \eqref{def_elp}, the vector fluctuations also enjoy
 {\it extra}  symmetries, that  shed light on the factorizability
 property of the vector equation \eqref{factpr1}.

\smallskip
 We introduce a new set of  operators
\be
P^{(\sigma)}_n\,=\,-e^{\frac{n}{4 M} \left( t+ r_\star 
\right)}\,\left[ 4 M \partial_t+n  \sigma \right]\,,
\ee
for an integer $n$. The quantity $\sigma$ in the previous definition is in principle arbitrary.
 These operators satisfy a Witt-like algebra
\bea
\[ P_m^{(\sigma_1)},  P_n^{(\sigma_2)} \]\,=\,(m-n)\,P_{m+n}^{(\sigma_3)}+m^2 (\sigma_1-\sigma_2)\,\delta_{m+n,0}\,,
 \eea
 with a contribution resembling a central extension. The quantity $\sigma_3$ is
 related to $\sigma_{1,2}$ by
\be
(m^2-n^2)\,\sigma_3\,=\,(m^2 \sigma_1-n^2 \sigma_2)(1-\delta_{m^2,n^2})\,.
\ee
When $\sigma_1=\sigma_2$, one finds the usual Witt algebra. The operators  $P_m^{(\sigma)}$ commute
with the equations of motion
\bea 
\label{seccom2}
\[{\cal E}_\ell, P_n^{(\sigma)}\]&=&0\,,
\eea
indicating they are also symmetries  of the system of equations.
 As for the commutation relations we met for the $L$'s operators, also
 the commutation relations for the $P$ operators of equation \eqref{seccom2} are
 valid everywhere in the exterior black hole geometry. As far as we are aware, we are the first to study the properties
 of these operators in a black hole space-time in modified gravity.
 
The associated quadratic Casimir gives
\be \label{sceq1a}
\ell (\ell+1) \beta(t,r)\,=\,P_0^2 [ \beta(t,r)]-\frac12 \left\{ P_1,P_{-1}\right\}
[ \beta(t,r)]\,.
\ee
The quantities $P_n^{(\sigma)}$ 
have also elegant commutation relations
with the operators $L_m$ defined in \eqref{def_elp}: we find
\bea
 (m+n)\,[ L_m, \,P_n^{(\sigma)}]&=&m^2\,L_{m+n}-n^2  P_{m+n}^{(\sigma)}
\,.
\eea
For $m+n=0$, the previous relation gives a trivial identity. To deal with this case, we introduce
another operator ($m\neq0$) 
\bea
 {\cal D}^{(\sigma)}&\equiv&\frac{2}{m}L_0-\frac{1}{m^2}
  [L_m, P_{-m}^{(\sigma)}]\,,\\
&=&\frac{\bar \Sigma}{\sqrt{\bar \Delta} }\,\partial_t \beta-\sqrt{\bar \Delta} \,\partial_r
 \beta-\sigma \beta
 \label{secexd}\,,
 \eea
which `closes the algebra' and commutes with all the remaining operators: 
\bea
0&=&
\[{\cal D}^{(\sigma_1)}, {\cal E}_\ell\]\,=\,\[{\cal D}^{(\sigma_1)}, P^\sigma_n\]\,=\,\[{\cal D}^{(\sigma_1)}, L_n\]
\,=\,\[{\cal D}^{(\sigma_1)}, {\cal D}^{(\sigma_2)}\]\,.
 \eea
Interestingly, comparing equations \eqref{def_lad} and \eqref{secexd}, we realize that for $\sigma = \sigma_\pm$
the operators ${\cal D}^{(\sigma_\pm)}$  coincide with the operators ${\cal D}^{\pm}$ of equation \eqref{def_lad}. We then find  an algebraic origin of the factorizable operators   ${\cal D}^{\pm}$ as related with a combination
of the two $SL(2, R)$ (or more generally, centerless Virasoro) algebras involving the operators $L_m$ and $P_n$. It would be interesting to find other examples of similar rich algebraic structures for other black holes in General Relativity
and beyond. 
   We now  demonstrate  very similar algebraic
structures for metric fluctuations, before turning to some
physical
implications of our findings  in section \ref{sec_pheno}.

\subsection{ Metric  perturbations}
\label{sec_met}

Interestingly, under particular hypotheses, symmetries identical  to  the ones we determined in the vector
sector of section \ref{sec_vec} also apply to the metric sector. Whilst, as we learned, the evolution 
equation for vector perturbations decouple from the remaining propagating degrees of freedom, metric 
fluctuations are sourced by vector fluctuations. To study the system more simply, we  choose appropriate
boundary conditions to set the independent vector fluctuations to zero, $\beta(t,r)=0$. We can then study the dynamics
of the metric perturbations only. We learned in equation \eqref{condH1} that the quantity $h_1(t,r)$ is algebraically
related with $h_0(t,r)$. Hence we focus on the dynamics of $h_0$: it is convenient to rescale  it 
as
\begin{equation}
h_0(t,r) = r^2 \gamma(t,r)\,.
\end{equation}
The metric perturbation $\gamma$ satisfies the evolution equation
\begin{equation}
\label{metr_eq1}
\mathcal{G}_\ell \,\left[ \gamma(t,r) \right]\, =\, 0\,,
\end{equation}
in the exterior geometry, with
\begin{equation}
\begin{split}
\label{defoGL}
\mathcal{G}_\ell \,\left[ \gamma(t,r)\right] & = \frac{\bar{\Sigma}(r)^2 \partial_t^2 \gamma}{\bar{\Delta}(r)} - \Bigg(\frac{2 \bar{\Sigma}(r)}{\sqrt{\bar{\Delta}(r)}} + \bar{\Sigma}'(r) \Bigg) \partial_t \gamma - 2\bar{\Sigma}(r) \partial_t \partial_r \gamma \\
& + 4\sqrt{\bar{\Delta}(r)} \, \partial_r \gamma  + \bar{\Delta}(r) \partial_r^2 \gamma
- (\ell-1)(\ell+2)\gamma\,.
\end{split}
\end{equation}
The functions $\bar \Delta$ and $\bar \Sigma$ are defined in the exterior region of the black hole in equations \eqref{ds1} and \eqref{dsi2}. 
Similarly to the vector case, the metric evolution equation \eqref{defoGL} may also be factorized into a product of two  first-order operators, with an identical structure.
\bea
\label{metfa1}
\label{factpr2a}
\mathcal{G}_\ell \,\left[ \gamma(t,r)\right] &=&\left( D^- \, D^+\right) \left[ \gamma(t,r)\right] \,=\,\left(D^+ \, D^-\right) \left[ \gamma(t,r)\right]\,, 
\eea
with
\bea
D^{\pm}\left[ \gamma(t,r)\right]&=&
\frac{\bar \Sigma}{\sqrt{\bar \Delta} }\,\partial_t \gamma-\sqrt{\bar \Delta} \,\partial_r
 \gamma-\rho_{\pm} \gamma\,.
 \label{def_ladme}
\eea
However this time the constants $\rho_\pm$ read
\bea
\rho_+&=&2+\ell \hskip0.5cm,\hskip0.5cm
\rho_-\,=\,1-\ell\,.
\eea
Hence the general solution to equation \eqref{metr_eq1} can be expressed as a linear
combination of solutions to equations $ D^- \left[ \gamma(t,r)\right]=0$ and  $ D^+ \left[ \gamma(t,r)\right]=0$. We find
the general
solution to be
\bea
 \gamma(t,r)&=&d_{+}\,\frac{S_{+}(t+r_\star)}{
(r+Q)^{\rho_{+}}}+d_{-}\,\frac{S_{-}(t+r_\star)}{
(r+Q)^{\rho_{-}}}\,,
\eea
with $d_\pm$ arbitrary constants, and  $S_{\pm}$ arbitrary functions of their argument.

\bigskip

Supplementing this, we find an underlying algebraic structure which lies behind the factorizability property of equation \eqref{metfa1}. It is actually {\it the very same structure} that applied to the vector case. In fact, the very same operators $L_n$, $P_m^\rho$ commute with the operator ${\cal G}_\ell$ 
introduced in equation \eqref{defoGL}, which controls the evolution equation for metric perturbations.
Consequently, the commutation relations remain the same as the
ones already studied in section \ref{sec_vec}. Combining these operators as in equation \eqref{secexd},
we can form the combinations ($m\neq0$)
\bea
 {\cal D}^{(\rho_\pm)}&\equiv&\left(\frac{2}{m}L_0-\frac{1}{m^2}
  [L_m, P_{-m}^{(\rho_\pm)}] \right)[\gamma(t,r)]\,,\\
&=&\frac{\bar \Sigma}{\sqrt{\bar \Delta} }\,\partial_t \beta-\sqrt{\bar \Delta} \,\partial_r
 \beta-\rho_\pm \beta
  \label{secexd2a}\,,
 \eea
which precisely coincide with the factorizing  operators  \eqref{def_ladme}. Equivalent to the vector fluctuations, in the case of the metric perturbations the structure of the evolution
equations can be related with $SL(2,R)$ symmetries of the system.

\section{Some consequences of the  conformal symmetries}
\label{sec_pheno}

The rich structure of symmetries associated with the evolution equations for
vector and metric fluctuations offers a deeper understanding
of the factorizability properties of the system, equations \eqref{factpr1} and \eqref{factpr2a},
which lead to exact time-dependent solutions for the system. Solutions
can be organized in $SL(2,R)$ multiplets, and have properties in common
with quasi-normal modes of black hole perturbations in General Relativity. There
are, however, some important differences. In this section, we seek to investigate these differences further.

\smallskip

We begin by focusing on the vector perturbations of section \ref{sec_vec}, since the metric
perturbations studied in section \ref{sec_met} behave very similarly. We consider
solutions belonging to the highest weight representation of the $SL(2,R)$
algebra associated with the operators $L_p$ (with $p=0,\pm1$).
We start with the time-dependent solution
corresponding to the highest weight (primary) vector. It is
  written
in equation \eqref{sol_bz}. We reiterate it  here substituting the values given in equation \eqref{def_sig}
for the quantities $\sigma_\pm$:
\be
\label{sol_bz1}
 \beta^{(0)}(t,r)\,=\,c_+\,
 \frac{e^{-\frac{(1+\ell)\,(t+r_\star)}{4 M}}}{ (r+Q)^{(1+\ell)}}+c_- \,{e^{\frac{\ell\,(t+r_\star)}{4 M}}}\,{ (r+Q)^{\ell}}\,.
\ee
The solution depends on two free parameters, $c_+$ and $c_-$; recall that the multipole
numbers satisfy $\ell\ge2$.  But the contribution to equation \eqref{sol_bz1} proportional to $c_-$ exponentially
grows in time. Due to this, it is not a physically interesting solution for describing  a small perturbation: hence we set $c_-=0$.  We are
left with a perturbation exponentially decaying in time, as expected for a black hole
quasi-normal mode (see e.g. \cite{Berti:2009kk} for a comprehensive review). However, the radial dependence
of the solution is somehow peculiar. Recall the definition of tortoise coordinate $r_\star$ of equation \eqref{def_tort}:
\bea
r_\star&=&\int^r \,\frac{\bar \Sigma(\tilde r)}{\bar \Delta (\tilde r)}\,d\tilde r
\,=\,\int^r 
\,\frac{\tilde r}{\tilde r+Q} \,\frac{\sqrt{Q^2+2 (M+Q)\tilde r}}{\tilde r-2M}\,d\tilde r\,.
\eea
In order for the integrand to be well defined for all $r>2M$, we impose the condition $(M+Q)\ge0$. 

For the case
$Q=-M$, the integral is particularly simple and gives  
\be
r_\star\,=\,2 M\,\ln{\left(\frac{r-2M}{\sqrt{r-M}} \right)}\,.
\ee
Hence we learn that $r_\star\,\sim\,2 M\,\ln{\left({r-2M} \right)}\to-\infty$ nearby the black hole horizon $r\to 2M$, while  $r_\star\,\sim\,2 M\,\ln{\left(r\right)}\to+\infty$ at large distances $r\to+\infty$ from the black hole. This is the expected behavior for the tortoise coordinate.  When expressed in terms
of the original $(t,r)$ coordinates,  the resulting exact solution for the highest weight vector is
\bea
\beta^{(0)}(t,r)\,=\,\frac{c_+\,e^{-\frac{(1+\ell)\,t}{4 M}}}{(r-2M)^{\frac{1+\ell}{2}}\,(r-M)^{\frac{3+3\ell}{4}}}\hskip0.7cm,\hskip0.7cm Q=-M\,.
\eea

 For $Q+M>0$ the integral can still be solved analytically, but the resulting formula is more complicated. It scales
as  $r_\star\,\sim\,\ln{\left({r-2M} \right)}$ near the black hole horizon, and  as $r_\star\,\sim\,r^{1/2}$ at plus infinity; again a behaviour compatible with what we should  expect from a tortoise coordinate. When expressed in terms
of the original $(t,r)$ coordinates,  the resulting exact solution for the highest weight vector is
\bea\beta^{(0)}(t,r)
&=&\frac{c_+\,e^{-\frac{(1+\ell)\,t}{4 M}}}{(r+Q)^{1+\ell}}
\,\left[\frac{2M+Q+\sqrt{Q^2+2 (M+Q)r}}{2M+Q-\sqrt{Q^2+2 (M+Q)r}} \right]^{\frac{1+\ell}{2}}\times
\nonumber
\\
&&\times e^{-\frac{1+\ell}{2 M}\,\sqrt{Q^2+2 (M+Q)r}}\,e^{\frac{(1+\ell)Q^2}{2 M \sqrt{Q (2 M+Q)}} {\rm arccot}\left( \sqrt{\frac{2MQ+Q^2}{Q^2+2 (M+Q) r}} \right)}\,\,.
\eea

\smallskip
For both cases, when $Q+M\ge0$
the resulting solution for the primary vector has a structure 
\be\label{primvec}
\beta^{(0)}(t,r)\,\propto\,
 \frac{e^{-\frac{(1+\ell)\,(t+r_\star)}{4 M}}}{ (r+Q)^{(1+\ell)}}\,.
\ee
It has the correct ingoing behaviour $e^{-i \omega(t+r_\star)}$ at the black hole horizon, $r_\star\to -\infty$, however, it does {\it not} exhibit 
the correct outgoing behavior $e^{-i \omega(t-r_\star)}$ at $r_\star\to +\infty$, as expected for a quasi-normal mode. The solution is better classified as a quasi-bound state, since there is energy dissipation inside the black hole horizon;
yet these modes decay exponentially at radial infinity. This behaviour is in common with perturbations of massive fields  \cite{Pani:2013pma,Dolan:2007mj,Rosa:2011my}. In our case, the non-minimal coupling of the vector with gravity induces contributions
to the energy momentum tensor of the vector fluctuations, that appear to mimic the effects of the vector mass, and cause
the aforementioned decay. 
 
 \medskip
 
 Let us now examine the behaviour of the frequencies of the solutions involved. The structure of the solution 
 \eqref{primvec} indicates that the highest weight primary vector has a purely imaginary frequency:
\be
 \omega_0\,=\,- i \frac{(1+\ell)}{4M}\,,
 \ee
with no real part.  
 Starting from the primary vector solution of equation \eqref{primvec}, we can generate its descendants $\beta^{(n)}$ by repeated application of the operators
$L_{-1}$, as explained around equation \eqref{def_desc}. The corresponding frequencies result
\be
\label{over_mg}
M\,\omega_n\,=\,M\,\omega_0-\frac{i\,n}{4}\hskip0.5cm,\hskip0.5cm n=0,1,2,3...
\ee
Hence the elements
of the highest weight representation have frequencies  shifted by an integer $n$ with respect to the frequency of the primary vector. 

For large values of $n$, this formula greatly resembles the behaviour
of frequency overtones of Schwarzschild black hole perturbations in General Relativity, which follows the law \cite{Padmanabhan:2003fx,Medved:2003rga}
\be
\label{over_gr}
{\text{GR, large $n$:}} \hskip0.6cm M \omega_n\,=\,\frac{\ln 3}{8 \pi}-\frac{i}{4} \left(n+\frac12 \right)+{\cal O}\left[ 
(n+1)^{-1/2}\right]
\ee
The relations between overtones and elements of a highest representation 
 were already noticed in the context of General Relativity \cite{Chen:2010ik}, exploiting near-horizon conformal symmetries
in proximity of the horizon. In our modified gravity set-up, the result extends in the entire exterior geometry, since our
conformal symmetries are defined in all exterior space. Comparing equations \eqref{over_mg} 
and \eqref{over_gr}, we notice that in 
 our modified gravity framework the frequencies are purely imaginary, and lack a real part. Moreover, as
explained above, we can not  talk of quasi-normal modes  in our system, but instead  of quasi-bound states. \\

\medskip

We have largely omitted discussion of the static solution, however we make a brief note here as a basis for further work. If one takes the zero-frequency limit of the time-dependent solutions considered in section \ref{sec_fluct}, one may directly solve the equations of motion in the static limit. With this procedure, the vector equation appears to become exactly that of the equation controlling the dynamics of perturbations around an extremal Reissner-Nordstr\"om configuration derived in \cite{Berens:2022ebl}
\begin{equation}
\bar{\Delta}(r) \, \partial_r^2 \beta(r) + 2\sqrt{\bar{\Delta}(r)} \, \partial_r \beta(r) - \ell(\ell+1)\beta(r) = 0
\end{equation}
where solutions may be obtained by repeated applications of the $L_{-1}$ operator to the highest weight vector, (or, conversely, applying the $L_{1}$ operator to the lowest weight vector) \cite{Charalambous:2022rre}. These solutions read
\be
\beta(r)\,=\,
 \frac{d_{(+)}}{ (r+Q)^{(1+\ell)}}+d_{(-)} \,{ (r+Q)^{\ell}}\,,
\ee
for two constants $d_{(\pm)}$. When $(Q+2M)>0$, both the contributions proportional to $d_{(+)}$ and $d_{(-)}$
are regular at the black hole horizon: subsequently, the associated Love numbers do not necessarily vanish. This highlights  
an immediate difference when compared to General Relativity, where typically only one among the two solutions of the second-order
static equation is physically acceptable.  If instead $(Q+2M)\le0$, then the solution proportional to $d_{(+)}$ is not regular
at the horizon, and should be discarded. Some care must be taken when considering the metric perturbations, as $h_1$ is algebraically connected to $h_0$ via equation \ref{condH1}, it would appear as if these perturbations immediately diverge on approach to the horizon. This divergence, however, may be removed (as usual) by moving into the tortoise coordinate, $r_*$, suggesting to us that this is purely a coordinate singularity.

The exterior background configuration that formed the primary focus of this work can be smoothly joined into a regular interior, and consequently
static solutions of vector and metric fluctuations may also be smoothly connected
with fluctuations in the interior \cite{Tasinato:2022vop} by the imposition of particular constraints on the relevant parameters. Due to our conformal symmetries appearing to be a global phenomena, it would be interesting to develop
our conformal symmetry structures into the interior solutions.

\section{Summary and Outlook}
\label{sec_out}

We studied the evolution of parity-odd, time-dependent field fluctuations  around an asymptotically flat solution
of a vector-tensor
theory of gravity. The geometry corresponds to a Schwarzschild black hole. However the dynamics of  parity-odd fluctuations -- both in the vector and in the metric sectors --  resemble the behaviour of perturbations 
around extremal black holes in General Relativity.  Their evolution equations, controlled by operators containing second order derivatives along time and space, can be factorized into a product of two commuting first order operators. This property allows us to analytically determine the most general time-dependent solutions  for the system of equations.  

We shown that the aforementioned factorizability property is associated with the existence of a large set of symmetries behind the system of equations. We identified two sets of operators belonging to  $SL(2,R)$ algebras, which represent  conformal symmetries for the system. Once combined, they  produce the aforementioned first-order operators that generate the equations of motion. Interestingly, while similar   conformal 
$SL(2,R)$ symmetries have been found in proximity of the horizon of a Schwarzschild configuration \cite{Bertini:2011ga}, our symmetries apply to  the entire exterior geometry of the black hole. 

We then studied the highest weight representation for one of the  $SL(2,R)$ algebras.
We analytically determined the expression for the highest weight   time-dependent solutions, and their  descendants. The highest weight representations  are known to have common properties with quasi-normal
modes of black holes in General Relativity. In fact, we found that elements belonging to the highest weight multiplets have frequencies separated by   integer numbers, resembling the behaviour of overtones of black hole quasi-normal modes. However, although our  solutions have the time-dependent profile of modes  ingoing into the black hole horizon, they do not describe   outgoing modes asymptotically far from it, as quasi-normal modes do. Instead, they  decay exponentially with the radial distance from the black hole, behaving as quasi-bound
states.

\smallskip
Our results leave many open questions for future work. We do not truly understand the origin
of our conformal symmetries, since we do not have  AdS asymptotic boundary conditions, nor do we consider configurations corresponding to  an extremal black hole. 
It would be nice to relate our conformal symmetries with some hidden symmetry of our vector-tensor system, or, at least, with   classes of its solutions within some
 specific Ansatz. A hint of this possibility was  discussed in section \ref{sec_setup} (building on \cite{Tasinato:2022vop}), where we pointed out that  our system admits  spherically symmetric  solutions with scaling symmetries.  Such solutions
 can be used to smooth out the black hole horizon, and describe ultracompact stars with a Schwarzschild exterior geometry. It may also be possible
that extra hidden symmetries relate our configurations with an asymptotically AdS space
in terms of subtracted  geometries -- as
shown in \cite{Bertini:2011ga} for Schwarzschild geometries within General Relativity --  or with some extremal or near-horizon black hole space-times,  hence geometrically motivating our conformal symmetries as some form of isometries.

\smallskip
As stated above, among the general time-dependent solutions for
vector and metric perturbations, the  highest weight
solutions are somehow special, and describe 
  quasi-bound states around the black hole geometry. It would be interesting to relate such  states with modes accounting for the black hole entropy, or the entropy
  of  regularized compact stars discussed in  section \ref{sec_setup} (see \cite{Tasinato:2022vop}).
  Regarding this important point, we have been able to extend $SL(2,R)$ symmetries to centerless Virasoro algebras.
 It would be interesting to
 investigate whether our symmetries can be formed in terms of center-full Virasoro symmetries, and
  find ways to include central charges. In fact,  the latter can be associated with the entropy of our geometrical configurations, analogous
 to the case for the Kerr black hole \cite{Castro:2010fd}. 

\smallskip
In this work, we focused exclusively on time-dependent solutions. However,  the system also admits static solutions, 
which can be determined  by  directly solving the static limit of the equations of motion, {\it or}
in terms of descendant (or ascendant) elements of  highest (lowest) weight representations \cite{Charalambous:2022rre} of the $SL(2,R)$ symmetries. By imposing appropriate boundary conditions, the static
solutions can be related with the Love numbers of our configurations, a topic which
received much attention in the recent literature in the context of emergent symmetries \cite{Hui:2021vcv,Charalambous:2022rre}. Some preliminary results
on Love numbers for our configurations have been explored in \cite{Tasinato:2022vop}, but certainly a more complete analysis is needed. 

\smallskip
We hope to be able to answer some of these  questions in the near future.

\subsection*{Acknowledgments}
We are
 partially funded by the STFC grants ST/T000813/1
and ST/X000648/1. For the purpose of open access, the authors have applied a Creative Commons Attribution licence to any Author Accepted Manuscript version arising.

\newpage

\begin{appendix}
\section{Evolution equations}
\label{appA_ev}
We collect in this appendix the three evolution equations for the parity-odd fluctuations $h_0$, $h_1$
and $\beta$, obtained varying the Lagrangian \eqref{starLAG} expanded at second order in perturbation
around the exterior solution of equations \eqref{extgeo1}-\eqref{extPI}:

 \begin{equation}
\begin{split}
0 & = 2r^4(Q+r)^2 \, \partial_t^2 \beta(t,r) - r^4 \sqrt{Q^2+2(M+Q)r} \, \partial_t \partial_r h_0(t,r) \\
& + 2r^2(2M-r)\Big(2Q \, \partial_r h_0(t,r) + (r-Q) \, \partial_t h_1(t,r)\Big) + 2r^3 \sqrt{Q^2 + 2(M+Q)r} \, \partial_t h_0(t,r) \\
& + r^3(r-2M)(Q+r)\, \partial_r^2 h_0(t,r) + r^4 \sqrt{Q^2 + 2(M+Q)r} \, \partial_t^2 h_1(t,r) \\
& + r^3(2M-r)(Q+r) \, \partial_t \partial_r h_1(t,r) - 2r^2(r-2M)^2(Q+r)^2 \, \partial_r^2 \beta(t,r) \\
& + 4r(2M-r)(Q+r)\Big(3MQ - (M+Q)r + r^2\Big) \, \partial_r \beta(t,r)\\
& - 2(2M-r)\Big(6MQ^2 + (\ell-1)(\ell+2)Q^2r + \ell(\ell+1)(2Qr^2 + r^3) - 2Mr^2 \Big)\beta(t,r) \\
& - (\ell-1)(\ell+2)(2M-r)r\sqrt{Q^2 + 2Mr + 2Qr}h_1(t,r) \\
& - \Bigg( 4Mr(r-3Q)-r\Big(\big(\ell(\ell+1)-8\big)Q + \ell(\ell+1)r\Big)\Bigg)h_0(t,r) \,,\\
\end{split}
\end{equation}
\begin{equation}
\begin{split}
0 & = \frac{2(\ell-1)(\ell+2)\sqrt{Q^2 + 2(M+Q)r}}{(Q+r)^2(r-2M)r} h_1(t,r) - \frac{2(\ell-1)(\ell+2)(Q^2+2(M+Q)r)}{(Q+r)^3(r-2M)^2}h_0(t,r)\\
& + \Big(\frac{2}{(Q+r)^2} - \frac{(\ell-1)(\ell+2)}{(r-2M)r} \Big) \beta(t,r) - \frac{2}{Q+r} \partial_r \beta(t,r) - \partial_r^2 \beta(t,r) \\
& + \frac{r\sqrt{Q^2 + 2(M+Q)r} }{(r-2M)(Q+r)} \partial_t \partial_r \beta(t,r)\\
& -\frac{(2MQ^3 + 6MQ(M+2Q)r + (14M^2 + 13MQ - 3Q^2)r^2 - 5(M+Q)r^3) }{(r-2M)^2(Q+r)^2\sqrt{Q^2 + 2(M+Q)r}}\partial_t \beta(t,r)  
 \,,
\\
\end{split}
\end{equation}
\begin{equation}
\begin{split}
0 = & \frac{r \sqrt{Q^2 + 2(M+Q)r}}{(r-2M)(Q+r)}\partial_t^2 \beta(t,r) + \partial_t \partial_r \beta(t,r) - \frac{1}{Q+r} \partial_t \beta(t,r) \\
& - \frac{(\ell-1)(\ell+2)\sqrt{Q^2 + 2(M+Q)r}}{r^2(Q+r)}\beta(t,r) - \frac{2(\ell-1)(\ell+2)}{r^2(Q+r)} h_1(t,r) \\
& - \frac{2(\ell-1)(\ell+2)\sqrt{Q^2 + 2(M+Q)r})}{r(2M-r)(Q+r)^2} h_0(t,r)
 \,.
\\
\label{app3}
\end{split}
\end{equation}
As explained in the main text, we can use \eqref{app3} to algebraically solve for $h_1$ as a function 
of the remaining variables.

\end{appendix}




\begin{thebibliography}{10}

\bibitem{Strominger:1997eq}
A.~Strominger, ``{Black hole entropy from near horizon microstates},''
  \href{http://dx.doi.org/10.1088/1126-6708/1998/02/009}{{\em JHEP} {\bfseries
  02} (1998) 009}, \href{http://arxiv.org/abs/hep-th/9712251}{{\ttfamily
  arXiv:hep-th/9712251}}.

\bibitem{Guica:2008mu}
M.~Guica, T.~Hartman, W.~Song, and A.~Strominger, ``{The Kerr/CFT
  Correspondence},'' \href{http://dx.doi.org/10.1103/PhysRevD.80.124008}{{\em
  Phys. Rev. D} {\bfseries 80} (2009) 124008},
  \href{http://arxiv.org/abs/0809.4266}{{\ttfamily arXiv:0809.4266 [hep-th]}}.

\bibitem{Bardeen:1999px}
J.~M. Bardeen and G.~T. Horowitz, ``{The Extreme Kerr throat geometry: A Vacuum
  analog of AdS(2) x S**2},''
  \href{http://dx.doi.org/10.1103/PhysRevD.60.104030}{{\em Phys. Rev. D}
  {\bfseries 60} (1999) 104030},
  \href{http://arxiv.org/abs/hep-th/9905099}{{\ttfamily arXiv:hep-th/9905099}}.

\bibitem{Govindarajan:2000ag}
T.~R. Govindarajan, V.~Suneeta, and S.~Vaidya, ``{Horizon states for AdS black
  holes},'' \href{http://dx.doi.org/10.1016/S0550-3213(00)00336-9}{{\em Nucl.
  Phys. B} {\bfseries 583} (2000) 291--303},
  \href{http://arxiv.org/abs/hep-th/0002036}{{\ttfamily arXiv:hep-th/0002036}}.

\bibitem{Birmingham:2001qa}
D.~Birmingham, K.~S. Gupta, and S.~Sen, ``{Near horizon conformal structure of
  black holes},'' \href{http://dx.doi.org/10.1016/S0370-2693(01)00354-9}{{\em
  Phys. Lett. B} {\bfseries 505} (2001) 191--196},
  \href{http://arxiv.org/abs/hep-th/0102051}{{\ttfamily arXiv:hep-th/0102051}}.

\bibitem{Hartman:2008pb}
T.~Hartman, K.~Murata, T.~Nishioka, and A.~Strominger, ``{CFT Duals for Extreme
  Black Holes},'' \href{http://dx.doi.org/10.1088/1126-6708/2009/04/019}{{\em
  JHEP} {\bfseries 04} (2009) 019},
  \href{http://arxiv.org/abs/0811.4393}{{\ttfamily arXiv:0811.4393 [hep-th]}}.

\bibitem{Carlip:1998wz}
S.~Carlip, ``{Black hole entropy from conformal field theory in any
  dimension},'' \href{http://dx.doi.org/10.1103/PhysRevLett.82.2828}{{\em Phys.
  Rev. Lett.} {\bfseries 82} (1999) 2828--2831},
  \href{http://arxiv.org/abs/hep-th/9812013}{{\ttfamily arXiv:hep-th/9812013}}.

\bibitem{Solodukhin:1998tc}
S.~N. Solodukhin, ``{Conformal description of horizon's states},''
  \href{http://dx.doi.org/10.1016/S0370-2693(99)00398-6}{{\em Phys. Lett. B}
  {\bfseries 454} (1999) 213--222},
  \href{http://arxiv.org/abs/hep-th/9812056}{{\ttfamily arXiv:hep-th/9812056}}.

\bibitem{Sachs:2001qb}
I.~Sachs and S.~N. Solodukhin, ``{Horizon holography},''
  \href{http://dx.doi.org/10.1103/PhysRevD.64.124023}{{\em Phys. Rev. D}
  {\bfseries 64} (2001) 124023},
  \href{http://arxiv.org/abs/hep-th/0107173}{{\ttfamily arXiv:hep-th/0107173}}.

\bibitem{Bertini:2011ga}
S.~Bertini, S.~L. Cacciatori, and D.~Klemm, ``{Conformal structure of the
  Schwarzschild black hole},''
  \href{http://dx.doi.org/10.1103/PhysRevD.85.064018}{{\em Phys. Rev. D}
  {\bfseries 85} (2012) 064018},
  \href{http://arxiv.org/abs/1106.0999}{{\ttfamily arXiv:1106.0999 [hep-th]}}.

\bibitem{Chen:2010ik}
B.~Chen and J.~Long, ``{Hidden Conformal Symmetry and Quasi-normal Modes},''
  \href{http://dx.doi.org/10.1103/PhysRevD.82.126013}{{\em Phys. Rev. D}
  {\bfseries 82} (2010) 126013},
  \href{http://arxiv.org/abs/1009.1010}{{\ttfamily arXiv:1009.1010 [hep-th]}}.

\bibitem{Castro:2010fd}
A.~Castro, A.~Maloney, and A.~Strominger, ``{Hidden Conformal Symmetry of the
  Kerr Black Hole},'' \href{http://dx.doi.org/10.1103/PhysRevD.82.024008}{{\em
  Phys. Rev. D} {\bfseries 82} (2010) 024008},
  \href{http://arxiv.org/abs/1004.0996}{{\ttfamily arXiv:1004.0996 [hep-th]}}.

\bibitem{Cvetic:2011hp}
M.~Cvetic and F.~Larsen, ``{Conformal Symmetry for General Black Holes},''
  \href{http://dx.doi.org/10.1007/JHEP02(2012)122}{{\em JHEP} {\bfseries 02}
  (2012) 122}, \href{http://arxiv.org/abs/1106.3341}{{\ttfamily arXiv:1106.3341
  [hep-th]}}.

\bibitem{Franzin:2011wi}
E.~Franzin and I.~Smolic, ``{A New look at hidden conformal symmetries of black
  holes},'' \href{http://dx.doi.org/10.1007/JHEP09(2011)081}{{\em JHEP}
  {\bfseries 09} (2011) 081}, \href{http://arxiv.org/abs/1107.2756}{{\ttfamily
  arXiv:1107.2756 [hep-th]}}.

\bibitem{Virmani:2012kw}
A.~Virmani, ``{Subtracted Geometry From Harrison Transformations},''
  \href{http://dx.doi.org/10.1007/JHEP07(2012)086}{{\em JHEP} {\bfseries 07}
  (2012) 086}, \href{http://arxiv.org/abs/1203.5088}{{\ttfamily arXiv:1203.5088
  [hep-th]}}.

\bibitem{Kim:2012mh}
Y.-W. Kim, Y.~S. Myung, and Y.-J. Park, ``{Quasinormal modes and hidden
  conformal symmetry in the Reissner-Nordstr\"om black hole},''
  \href{http://dx.doi.org/10.1140/epjc/s10052-013-2440-8}{{\em Eur. Phys. J. C}
  {\bfseries 73} (2013) 2440}, \href{http://arxiv.org/abs/1205.3701}{{\ttfamily
  arXiv:1205.3701 [hep-th]}}.

\bibitem{Ghezelbash:2012qn}
A.~M. Ghezelbash and H.~M. Siahaan, ``{Hidden and Generalized Conformal
  Symmetry of Kerr-Sen Spacetimes},''
  \href{http://dx.doi.org/10.1088/0264-9381/30/13/135005}{{\em Class. Quant.
  Grav.} {\bfseries 30} (2013) 135005},
  \href{http://arxiv.org/abs/1206.0714}{{\ttfamily arXiv:1206.0714 [hep-th]}}.

\bibitem{Charalambous:2022rre}
P.~Charalambous, S.~Dubovsky, and M.~M. Ivanov, ``{Love symmetry},''
  \href{http://dx.doi.org/10.1007/JHEP10(2022)175}{{\em JHEP} {\bfseries 10}
  (2022) 175}, \href{http://arxiv.org/abs/2209.02091}{{\ttfamily
  arXiv:2209.02091 [hep-th]}}.

\bibitem{EventHorizonTelescope:2019dse}
{\bfseries Event Horizon Telescope} Collaboration, K.~Akiyama {\em et~al.},
  ``{First M87 Event Horizon Telescope Results. I. The Shadow of the
  Supermassive Black Hole},''
  \href{http://dx.doi.org/10.3847/2041-8213/ab0ec7}{{\em Astrophys. J. Lett.}
  {\bfseries 875} (2019) L1}, \href{http://arxiv.org/abs/1906.11238}{{\ttfamily
  arXiv:1906.11238 [astro-ph.GA]}}.

\bibitem{Gralla:2019xty}
S.~E. Gralla, D.~E. Holz, and R.~M. Wald, ``{Black Hole Shadows, Photon Rings,
  and Lensing Rings},''
  \href{http://dx.doi.org/10.1103/PhysRevD.100.024018}{{\em Phys. Rev. D}
  {\bfseries 100} no.~2, (2019) 024018},
  \href{http://arxiv.org/abs/1906.00873}{{\ttfamily arXiv:1906.00873
  [astro-ph.HE]}}.

\bibitem{Johnson:2019ljv}
M.~D. Johnson {\em et~al.}, ``{Universal interferometric signatures of a black
  hole\textquoteright{}s photon ring},''
  \href{http://dx.doi.org/10.1126/sciadv.aaz1310}{{\em Sci. Adv.} {\bfseries 6}
  no.~12, (2020) eaaz1310}, \href{http://arxiv.org/abs/1907.04329}{{\ttfamily
  arXiv:1907.04329 [astro-ph.IM]}}.

\bibitem{Himwich:2020msm}
E.~Himwich, M.~D. Johnson, A.~Lupsasca, and A.~Strominger, ``{Universal
  polarimetric signatures of the black hole photon ring},''
  \href{http://dx.doi.org/10.1103/PhysRevD.101.084020}{{\em Phys. Rev. D}
  {\bfseries 101} no.~8, (2020) 084020},
  \href{http://arxiv.org/abs/2001.08750}{{\ttfamily arXiv:2001.08750 [gr-qc]}}.

\bibitem{Gralla:2020yvo}
S.~E. Gralla and A.~Lupsasca, ``{Observable shape of black hole photon
  rings},'' \href{http://dx.doi.org/10.1103/PhysRevD.102.124003}{{\em Phys.
  Rev. D} {\bfseries 102} no.~12, (2020) 124003},
  \href{http://arxiv.org/abs/2007.10336}{{\ttfamily arXiv:2007.10336 [gr-qc]}}.

\bibitem{Gralla:2020srx}
S.~E. Gralla, A.~Lupsasca, and D.~P. Marrone, ``{The shape of the black hole
  photon ring: A precise test of strong-field general relativity},''
  \href{http://dx.doi.org/10.1103/PhysRevD.102.124004}{{\em Phys. Rev. D}
  {\bfseries 102} no.~12, (2020) 124004},
  \href{http://arxiv.org/abs/2008.03879}{{\ttfamily arXiv:2008.03879 [gr-qc]}}.

\bibitem{Li:2021zct}
P.-C. Li, T.-C. Lee, M.~Guo, and B.~Chen, ``{Correspondence of eikonal
  quasinormal modes and unstable fundamental photon orbits for a Kerr-Newman
  black hole},'' \href{http://dx.doi.org/10.1103/PhysRevD.104.084044}{{\em
  Phys. Rev. D} {\bfseries 104} no.~8, (2021) 084044},
  \href{http://arxiv.org/abs/2105.14268}{{\ttfamily arXiv:2105.14268 [gr-qc]}}.

\bibitem{Raffaelli:2021gzh}
B.~Raffaelli, ``{Hidden conformal symmetry on the black hole photon sphere},''
  \href{http://dx.doi.org/10.1007/JHEP03(2022)125}{{\em JHEP} {\bfseries 03}
  (2022) 125}, \href{http://arxiv.org/abs/2112.12543}{{\ttfamily
  arXiv:2112.12543 [gr-qc]}}.

\bibitem{Hadar:2022xag}
S.~Hadar, D.~Kapec, A.~Lupsasca, and A.~Strominger, ``{Holography of the photon
  ring},'' \href{http://dx.doi.org/10.1088/1361-6382/ac8d43}{{\em Class. Quant.
  Grav.} {\bfseries 39} no.~21, (2022) 215001},
  \href{http://arxiv.org/abs/2205.05064}{{\ttfamily arXiv:2205.05064 [gr-qc]}}.

\bibitem{Kapec:2022dvc}
D.~Kapec, A.~Lupsasca, and A.~Strominger, ``{Photon rings around warped black
  holes},'' \href{http://dx.doi.org/10.1088/1361-6382/acc164}{{\em Class.
  Quant. Grav.} {\bfseries 40} no.~9, (2023) 095006},
  \href{http://arxiv.org/abs/2211.01674}{{\ttfamily arXiv:2211.01674 [gr-qc]}}.

\bibitem{Barack:2018yly}
L.~Barack {\em et~al.}, ``{Black holes, gravitational waves and fundamental
  physics: a roadmap},'' \href{http://dx.doi.org/10.1088/1361-6382/ab0587}{{\em
  Class. Quant. Grav.} {\bfseries 36} no.~14, (2019) 143001},
  \href{http://arxiv.org/abs/1806.05195}{{\ttfamily arXiv:1806.05195 [gr-qc]}}.

\bibitem{Tasinato:2014eka}
G.~Tasinato, ``{Cosmic Acceleration from Abelian Symmetry Breaking},''
  \href{http://dx.doi.org/10.1007/JHEP04(2014)067}{{\em JHEP} {\bfseries 04}
  (2014) 067}, \href{http://arxiv.org/abs/1402.6450}{{\ttfamily arXiv:1402.6450
  [hep-th]}}.

\bibitem{Heisenberg:2014rta}
L.~Heisenberg, ``{Generalization of the Proca Action},''
  \href{http://dx.doi.org/10.1088/1475-7516/2014/05/015}{{\em JCAP} {\bfseries
  05} (2014) 015}, \href{http://arxiv.org/abs/1402.7026}{{\ttfamily
  arXiv:1402.7026 [hep-th]}}.

\bibitem{Chagoya:2016aar}
J.~Chagoya, G.~Niz, and G.~Tasinato, ``{Black Holes and Abelian Symmetry
  Breaking},'' \href{http://dx.doi.org/10.1088/0264-9381/33/17/175007}{{\em
  Class. Quant. Grav.} {\bfseries 33} no.~17, (2016) 175007},
  \href{http://arxiv.org/abs/1602.08697}{{\ttfamily arXiv:1602.08697
  [hep-th]}}.

\bibitem{Tasinato:2022vop}
G.~Tasinato, ``{Ultracompact vector stars},''
  \href{http://dx.doi.org/10.1103/PhysRevD.106.044022}{{\em Phys. Rev. D}
  {\bfseries 106} no.~4, (2022) 044022},
  \href{http://arxiv.org/abs/2205.05311}{{\ttfamily arXiv:2205.05311 [gr-qc]}}.

\bibitem{Minamitsuji:2016ydr}
M.~Minamitsuji, ``{Solutions in the generalized Proca theory with the
  nonminimal coupling to the Einstein tensor},''
  \href{http://dx.doi.org/10.1103/PhysRevD.94.084039}{{\em Phys. Rev. D}
  {\bfseries 94} no.~8, (2016) 084039},
  \href{http://arxiv.org/abs/1607.06278}{{\ttfamily arXiv:1607.06278 [gr-qc]}}.

\bibitem{Cisterna:2016nwq}
A.~Cisterna, M.~Hassaine, J.~Oliva, and M.~Rinaldi, ``{Static and rotating
  solutions for Vector-Galileon theories},''
  \href{http://dx.doi.org/10.1103/PhysRevD.94.104039}{{\em Phys. Rev. D}
  {\bfseries 94} no.~10, (2016) 104039},
  \href{http://arxiv.org/abs/1609.03430}{{\ttfamily arXiv:1609.03430 [gr-qc]}}.

\bibitem{Babichev:2017rti}
E.~Babichev, C.~Charmousis, and M.~Hassaine, ``{Black holes and solitons in an
  extended Proca theory},''
  \href{http://dx.doi.org/10.1007/JHEP05(2017)114}{{\em JHEP} {\bfseries 05}
  (2017) 114}, \href{http://arxiv.org/abs/1703.07676}{{\ttfamily
  arXiv:1703.07676 [gr-qc]}}.

\bibitem{Chagoya:2017fyl}
J.~Chagoya, G.~Niz, and G.~Tasinato, ``{Black Holes and Neutron Stars in Vector
  Galileons},'' \href{http://dx.doi.org/10.1088/1361-6382/aa7c01}{{\em Class.
  Quant. Grav.} {\bfseries 34} no.~16, (2017) 165002},
  \href{http://arxiv.org/abs/1703.09555}{{\ttfamily arXiv:1703.09555 [gr-qc]}}.

\bibitem{Heisenberg:2017xda}
L.~Heisenberg, R.~Kase, M.~Minamitsuji, and S.~Tsujikawa, ``{Hairy black-hole
  solutions in generalized Proca theories},''
  \href{http://dx.doi.org/10.1103/PhysRevD.96.084049}{{\em Phys. Rev. D}
  {\bfseries 96} no.~8, (2017) 084049},
  \href{http://arxiv.org/abs/1705.09662}{{\ttfamily arXiv:1705.09662 [gr-qc]}}.

\bibitem{Chagoya:2017ojn}
J.~Chagoya and G.~Tasinato, ``{Stealth configurations in vector-tensor theories
  of gravity},'' \href{http://dx.doi.org/10.1088/1475-7516/2018/01/046}{{\em
  JCAP} {\bfseries 01} (2018) 046},
  \href{http://arxiv.org/abs/1707.07951}{{\ttfamily arXiv:1707.07951
  [hep-th]}}.

\bibitem{Filippini:2017kov}
F.~Filippini and G.~Tasinato, ``{An exact solution for a rotating black hole in
  modified gravity},''
  \href{http://dx.doi.org/10.1088/1475-7516/2018/01/033}{{\em JCAP} {\bfseries
  01} (2018) 033}, \href{http://arxiv.org/abs/1709.02147}{{\ttfamily
  arXiv:1709.02147 [hep-th]}}.

\bibitem{Kase:2018voo}
R.~Kase, M.~Minamitsuji, S.~Tsujikawa, and Y.-L. Zhang, ``{Black hole
  perturbations in vector-tensor theories: The odd-mode analysis},''
  \href{http://dx.doi.org/10.1088/1475-7516/2018/02/048}{{\em JCAP} {\bfseries
  02} (2018) 048}, \href{http://arxiv.org/abs/1801.01787}{{\ttfamily
  arXiv:1801.01787 [gr-qc]}}.

\bibitem{Berens:2022ebl}
R.~Berens and L.~Hui, ``{Ladder Symmetries of Black Holes and de Sitter Space:
  Love Numbers and Quasinormal Modes},''
  \href{http://arxiv.org/abs/2212.09367}{{\ttfamily arXiv:2212.09367
  [hep-th]}}.

\bibitem{Hui:2021vcv}
L.~Hui, A.~Joyce, R.~Penco, L.~Santoni, and A.~R. Solomon, ``{Ladder symmetries
  of black holes. Implications for love numbers and no-hair theorems},''
  \href{http://dx.doi.org/10.1088/1475-7516/2022/01/032}{{\em JCAP} {\bfseries
  01} no.~01, (2022) 032}, \href{http://arxiv.org/abs/2105.01069}{{\ttfamily
  arXiv:2105.01069 [hep-th]}}.

\bibitem{Cardoso:2017qmj}
V.~Cardoso, T.~Houri and M.~Kimura,
``{Mass Ladder Operators from Spacetime Conformal Symmetry},''
\href{doi:10.1103/PhysRevD.96.024044} {Phys. Rev. D \textbf{96} (2017) no.2, 024044}
\href{http://arxiv.org/abs/1706.07339}{{\ttfamily
  arXiv:1706.07339 [hep-th]}}.

\bibitem{BenAchour:2022uqo}
J.~Ben Achour, E.~R.~Livine, S.~Mukohyama and J.~P.~Uzan,
``{Hidden symmetry of the static response of black holes: applications to Love numbers,}''
\href{doi:10.1007/JHEP07(2022)112} {
JHEP \textbf{07} (2022), 112}
\href{http://arxiv.org/abs/2202.12828}{{\ttfamily
 arXiv:2202.12828 [gr-qc]}}.




\bibitem{Berti:2009kk}
E.~Berti, V.~Cardoso, and A.~O. Starinets, ``{Quasinormal modes of black holes
  and black branes},''
  \href{http://dx.doi.org/10.1088/0264-9381/26/16/163001}{{\em Class. Quant.
  Grav.} {\bfseries 26} (2009) 163001},
  \href{http://arxiv.org/abs/0905.2975}{{\ttfamily arXiv:0905.2975 [gr-qc]}}.

\bibitem{Pani:2013pma}
P.~Pani, ``{Advanced Methods in Black-Hole Perturbation Theory},''
  \href{http://dx.doi.org/10.1142/S0217751X13400186}{{\em Int. J. Mod. Phys. A}
  {\bfseries 28} (2013) 1340018},
  \href{http://arxiv.org/abs/1305.6759}{{\ttfamily arXiv:1305.6759 [gr-qc]}}.

\bibitem{Dolan:2007mj}
S.~R. Dolan, ``{Instability of the massive Klein-Gordon field on the Kerr
  spacetime},'' \href{http://dx.doi.org/10.1103/PhysRevD.76.084001}{{\em Phys.
  Rev. D} {\bfseries 76} (2007) 084001},
  \href{http://arxiv.org/abs/0705.2880}{{\ttfamily arXiv:0705.2880 [gr-qc]}}.

\bibitem{Rosa:2011my}
J.~G. Rosa and S.~R. Dolan, ``{Massive vector fields on the Schwarzschild
  spacetime: quasi-normal modes and bound states},''
  \href{http://dx.doi.org/10.1103/PhysRevD.85.044043}{{\em Phys. Rev. D}
  {\bfseries 85} (2012) 044043},
  \href{http://arxiv.org/abs/1110.4494}{{\ttfamily arXiv:1110.4494 [hep-th]}}.

\bibitem{Padmanabhan:2003fx}
T.~Padmanabhan, ``{Quasinormal modes: A Simple derivation of the level spacing
  of the frequencies},''
  \href{http://dx.doi.org/10.1088/0264-9381/21/1/L01}{{\em Class. Quant. Grav.}
  {\bfseries 21} (2004) L1},
  \href{http://arxiv.org/abs/gr-qc/0310027}{{\ttfamily arXiv:gr-qc/0310027}}.

\bibitem{Medved:2003rga}
A.~J.~M. Medved, D.~Martin, and M.~Visser, ``{Dirty black holes: Quasinormal
  modes},'' \href{http://dx.doi.org/10.1088/0264-9381/21/6/008}{{\em Class.
  Quant. Grav.} {\bfseries 21} (2004) 1393--1406},
  \href{http://arxiv.org/abs/gr-qc/0310009}{{\ttfamily arXiv:gr-qc/0310009}}.

\end{thebibliography}
\providecommand{\href}[2]{#2}\begingroup\raggedright\endgroup

\end{document}